\documentclass[epj]{svjour}

\usepackage{graphics}
\usepackage{epsfig}

\newcommand{\be}{\begin{equation}}
\newcommand{\ee}{\end{equation}}
\newcommand{\bc}{\begin{center}}
\newcommand{\ec}{\end{center}}
\newcommand{\bfig}{\begin{figure}}
\newcommand{\efig}{\end{figure}}
\newcommand{\bea}{\begin{eqnarray}}
\newcommand{\eea}{\end{eqnarray}}

\newcommand{\tpl}{t_{{\rm w}}}
\newcommand{\tlow}{t_{{\rm low}}}
\newcommand{\Tpl}{T_{{\rm pl}}}

\begin{document}

\title{Frequency Dependence of Aging, Rejuvenation and Memory in
a disordered ferroelectric}

\author{J.-Ph.
Bouchaud\inst{1}
\and  P. Doussineau\inst{2}
\and T. de Lacerda-Ar\^oso\inst{3}
\and A. Levelut \inst{2}}
\institute{Service de Physique de l'Etat Condens\'e, CEA Saclay,\\ 91191 Gif sur Yvette Cedex, France
\and
Laboratoire des Milieux D\'esordonn\'es et H\'et\'erog\`enes,\\
Universit\'e P. et M. Curie, Case 86, 75252 Paris Cedex 05, France.\\
(Associated with the Centre National de la Recherche Scientifique (UMR
7603)
\and
Departamento de Fisica, Universidade do Minho, 4709 Braga, Portugal.}
\date{\today}

\abstract{
We characterize in details the aging properties of
the ferroelectric phase of K Ta$_{1-x}$ Nb$_x$ O$_3$ ({\sc ktn}), where
both rejuvenation and (partial) memory are observed. In particular, we
carefully examine the frequency dependence of several quantities that
characterize aging, rejuvenation and memory. We find a marked subaging
behaviour, with an a.c. dielectric susceptiblity scaling as $\omega \sqrt{\tpl}$,
where $\tpl$ is the waiting time. We suggest an interpretation in
terms of pinned domain walls, much along the lines
proposed for aging in a disordered ferromagnet, where {\it both} domain wall 
reconformations and overall (cumulative) domain growth are needed to
rationalize the experimental findings.
}
\authorrunning{J.-P. Bouchaud \emph{et al.}}

\titlerunning{Frequency Dependence of Aging in
a disordered ferroelectric}
\maketitle

\section{Introduction}

Aging is a widespread phenomenom which manifests itself
through the dependence of the properties of a material on the history of the 
studied sample \cite{review}. It has been observed and widely documented on
the magnetic susceptibility of spin-glasses (SG) \cite{1}-\cite{4} and
disordered ferromagnets \cite{12}, on the elastic compliance of polymers
\cite{5}, on the dielectric constant of disordered dielectrics (`dipolar
glasses') \cite{6,7} and of some liquids \cite{8}, and recently on the
rheological properties and dynamical structure factor of `pasty' colloids \cite{colloids1,colloids2}. Aging reflects the fact
that the time needed for the system
to equilibrate becomes larger than the experimental time scales. Although
several theoretical models for aging have been investigated, the correct
physical picture for aging is still rather controversial, in particular in
spin-glasses. The simplest model for aging is that of domain growth: below
some critical temperature, the system progressively orders by growing domains
of the competing low temperature phases
(for example up and down in the case of an Ising ferromagnet). In this case,
the aging
of the susceptibility can be understood from the dynamics of pinned domain
walls: as time
elapses, the density of domain walls decreases, and the remaining domain
walls are
better and better pinned.

Stronger constraints on microscopic theories appear when one wants to
interpret the
so-called `rejuvenation and memory' effect \cite{rejmem}. Suppose that one
cools a glassy system at a constant cooling rate ${\cal R}$ down to a certain
temperature $\Tpl$, which is
then held constant. At this temperature, aging manifests itself by the
decrease of some susceptibility (both of its real and imaginary part). If
cooling is then resumed after a `plateau time' $\tpl$, the susceptibility is
seen to increase again, approaching the value  it would have had if $\tpl$
was zero: the system `rejuvenates' as it seems to forget its
stay at $\Tpl$. If cooling is carried further, down to low temperatures and
then followed by a steady heating, a `dip' in the susceptibility is observed
when passing back through the temperature $\Tpl$. This shows that the system
has actually kept some `memory' of its evolution at $\Tpl$. The coexistence
of these two effects is quite remarkable, and has led
to several theoretical scenarios \cite{Swedes,JP,Yoshino,Sasaki}. These two
effects have
been seen in a variety of materials: various spin-glasses, {\sc pmma}
(plexiglass) \cite{11} or {\sc ktn} (a disordered dielectric, see below)
\cite{13}. However, some materials exhibit aging
but no rejuvenation \cite{prl}, or rejuvenation but no (or partial) memory
\cite{12}.

The aim of this paper is to characterize in more details the aging properties
of
the ferroelectric phase of {\sc ktn}, where both rejuvenation and
(partial) memory
are observed \cite{13}. In particular, we have carefully examined the
frequency dependence of several quantities which characterize aging, 
rejuvenation and memory.
We suggest an interpretation in terms of pinned domain walls, much along the 
lines
proposed in \cite{12} for a disordered ferromagnet, where both domain wall
{\it reconformations} and overall domain growth are needed to rationalize the
experimental
findings.

\section{Experiments}

The pure potassium tantalate crystal K Ta O$_3$ belongs
to the cubic perovskite family. Quantum fluctuations prevents a ferroelectric
order to establish fully, even at $T=0$K.
If a fraction $x$ of tantalum ions is randomly substituted by isoelectronic
niobium
ions to give crystals of K Ta$_{1-x}$ Nb$_x$ O$_3$ ({\sc ktn}), the trend
towards ferroelectricity of the perovskite lattice is enhanced. It has been
assumed that this tendency was due to random fields generated by the niobium
ions \cite{16}. Indeed, if the niobium concentration $x$ is larger than
$x_c=0.008$, the cooperative ordering gives way to a ferroelectric phase at
temperatures below a finite transition temperature $T_c$. For
the sample used in the present study we observe a broad transition at
$T_c=31$ K, defined by the maxima of the real part $\epsilon'$ and the
imaginary part $\epsilon''$ of the dielectric constant, independently of the measuring angular frequency $\omega$.
When $x>x_c$, there are actually several transitions: when the temperature is
lowered,
{\sc ktn} is successively cubic, tetragonal, orthorhombic and finally,
rhombohedral \cite{17}. For the concentration $x$ of our sample, the three
transitions merge into a single one. From the phase diagram obtained in
\cite{17}, we infer that the niobium
concentration is around $x=0.022$. The experiments reported below were all 
performed at temperatures $T < 25$ K. Therefore, they concern an ordered
(ferroelectric) phase which contains some amount of disorder (the randomly
substituted niobium ions). 

Using a Hewlett-Packard 4192A impedance analyser,
we have measured the electric
capacitance and the dielectric loss at seven frequencies, ranging from $f=1$
kHz to $f=1$ MHz. These can easily be transformed into the real part
$\epsilon'$ and the imaginary part $\epsilon''$ of the complex dielectric
constant. We measure the complex capacitance $C(\omega,t,T)$ as a
function of time $t$ while the sample temperature $T$ is a controled function
of time. Several procedures were used after an annealing near $55$ K and an
initial rapid cooling across the transition temperature $T_c$ down to
$T_{\max}=22$ K. The simplest procedure was composed of a cooling at a
constant rate $dT/dt=-{\cal R}$ from $T_{\max}$ to
$T_{\min}=4.8$K, immediately followed by a steady heating at the opposite
rate
$dT/dt=+{\cal R}$ from $T_{\min}$ to $T_{\max}$: this was used as the {\it
reference curves}.  Another procedure is to cool at the same rate $-{\cal R}$
between $T_{\max}$ and an intermediate temperature $\Tpl$. There, cooling is
interrupted by an isothermal evolution or plateau at this temperature, which
lasts typically $\tpl = 10 000$ seconds. Cooling then resumes down to
$T_{\min}$ and is again immediately followed by a steady heating at the
opposite rate from $T_{\min}$ to $T_{\max}$. A typical cooling rate is ${\cal
R}=5.2$ mK/s. It was carefully checked that the linear response conditions are fulfilled.

\section{Frequency analysis of isothermal aging}

The isothermal variation of
the complex capacitance $C(\omega,t)$ in {\sc ktn} is a slow decay in time 
(which also depends on the cooling rate ${\cal R}$ \cite{18}). This decay can
be fitted by a power law $(t_0+t)^{-\eta}$ with a small exponent $\eta$
(typically $\simeq 0.05$) and $t_0$ in the range of $10^2$ s. 
For a given 
plateau temperature $\Tpl$ and a given plateau duration
$\tpl$, the total decay of the complex capacitance, $\delta C(\omega,\tpl,
\Tpl)\equiv C(\omega,t=0^+,\Tpl)-C(\omega,t=\tpl,\Tpl) \geq 0$ is found to
obey a negative power law of the frequency $f$ and therefore of $\omega=2\pi
f$. The real part and the imaginary part of this variation can be fitted by
(see Figure 1):
\be
\delta C'(\omega,\tpl,\Tpl)=N' \omega^{-\nu}   \qquad \delta
C''(\omega,\tpl,\Tpl)=N'' \omega^{-\nu},\label{lognu}
\ee
with the same exponent $\nu$. This is actually a general consequence of the
Kramers-Kr\"onig
relations, which hold for a linear, stationary and causal response theory.
Although aging
means that the response is non stationary, these Kramers-Kr\"onig relations
are valid
in the quasi stationary limit where $\omega^{-1} \ll \tpl$, which obviously
holds in our
experiments (where $\omega \tpl \gg 10^3$). The Kramers-Kr\"onig 
equations actually also
provide a precise relation between
the amplitudes
$N'$ and $N''$, which only depends on the value of $\nu$:
\be
\frac{N''}{N'}= \tan\left(\frac{\nu \pi}{2}\right).\label{tannu}
\ee

\bfig[t]
\bc
\epsfxsize=\linewidth
\epsfbox{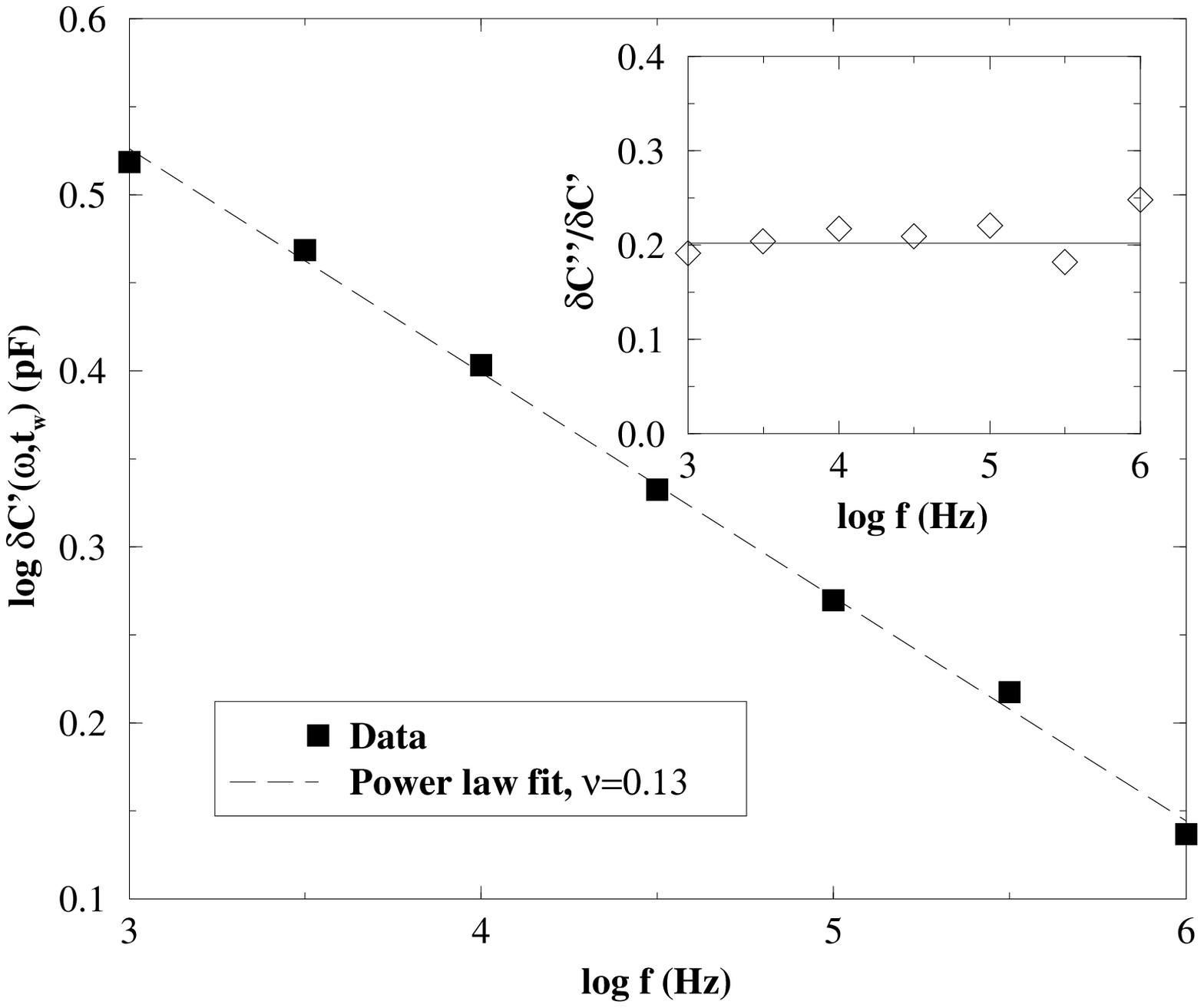}
\caption{Behaviour of the total decay of the real-part of the capacitance, 
$\delta C'(\omega,\tpl,
\Tpl)$ $\equiv C'(\omega,t=0^+,\Tpl)-C'(\omega,t=\tpl,\Tpl)$, as a function of the
frequency $f=\omega/2\pi$, for $\tpl=20 000$ s and $\Tpl=12.1$ K.
The magnitude of this decay is found to behave as a power-law
of the frequency: $\delta C' \propto \omega^{-\nu}$, with 
$\nu=0.127 \pm 0.005$ over
three decades in frequency ($1$ kHz -- $1$ MHz.). We show in the inset 
the ratio
$\delta C''/\delta C'$ as a function of frequency. For a power-law 
dependence, this ratio should be constant, equal to $\tan(\nu\pi/2)=0.202$,
shown as the full line.
\label{fig1}}
\ec
\efig

This relation provides a second determination of $\nu$, by checking the
constancy of the ratio of the imaginary and real parts of $\delta C$, and
using Eq. (\ref{tannu}) to extract $\nu$. As shown in Fig. 1, obtained for
$\Tpl=12.1$ K and $\tpl=20 000$ s, these two
procedures allow one to obtain rather close estimates for $\nu$. The
regression in log-scale based on Eq. (\ref{lognu}) gives $\nu =0.127 \pm
0.005$. The exponent $\nu$ is found to depend weakly both
on temperature and time. More precisely, it varies linearly with $\Tpl$ but only
logarithmically with $\tpl$. The magnitudes $N'$ and $N''$ are increasing
functions of $\Tpl$ and $\tpl$.

In spin glasses, the dependencies of the aging part of the susceptibility on 
the waiting time $\tpl$ and on the frequency $\omega$ are not independent,
but rather follow a simple
$\omega \tpl$ scaling, which means that the typical relaxation time after
waiting $\tpl$
is $\tpl$ itself. More generally, one may expect scaling with $\omega
\tpl^\mu$, where $\mu$ is an exponent, generally found to be smaller than
$1$: this corresponds to `sub-aging' \cite{review,JP}.
It is interesting to look for a similar scaling in our data. As shown in Fig. 2, the aging part of the capacitance can indeed be approximately rescaled
with $\mu \simeq 0.5$, leading to:
\be
C(\omega,\tpl,\Tpl) \simeq C_{\rm st}(\omega,\Tpl) + {\cal C}(\Tpl) \times (\omega
\tpl^\mu)^{-\nu} \quad \tpl \gg t_0,
\ee
where $C_{\rm st}$ is the stationary part of the response. This equation describes simultaneously both the above discussed frequency dependence and the time relaxation of
the aging part, with the exponent relation $\eta=\nu\mu$.

\bfig[t]
\bc
\epsfxsize=\linewidth
\epsfbox{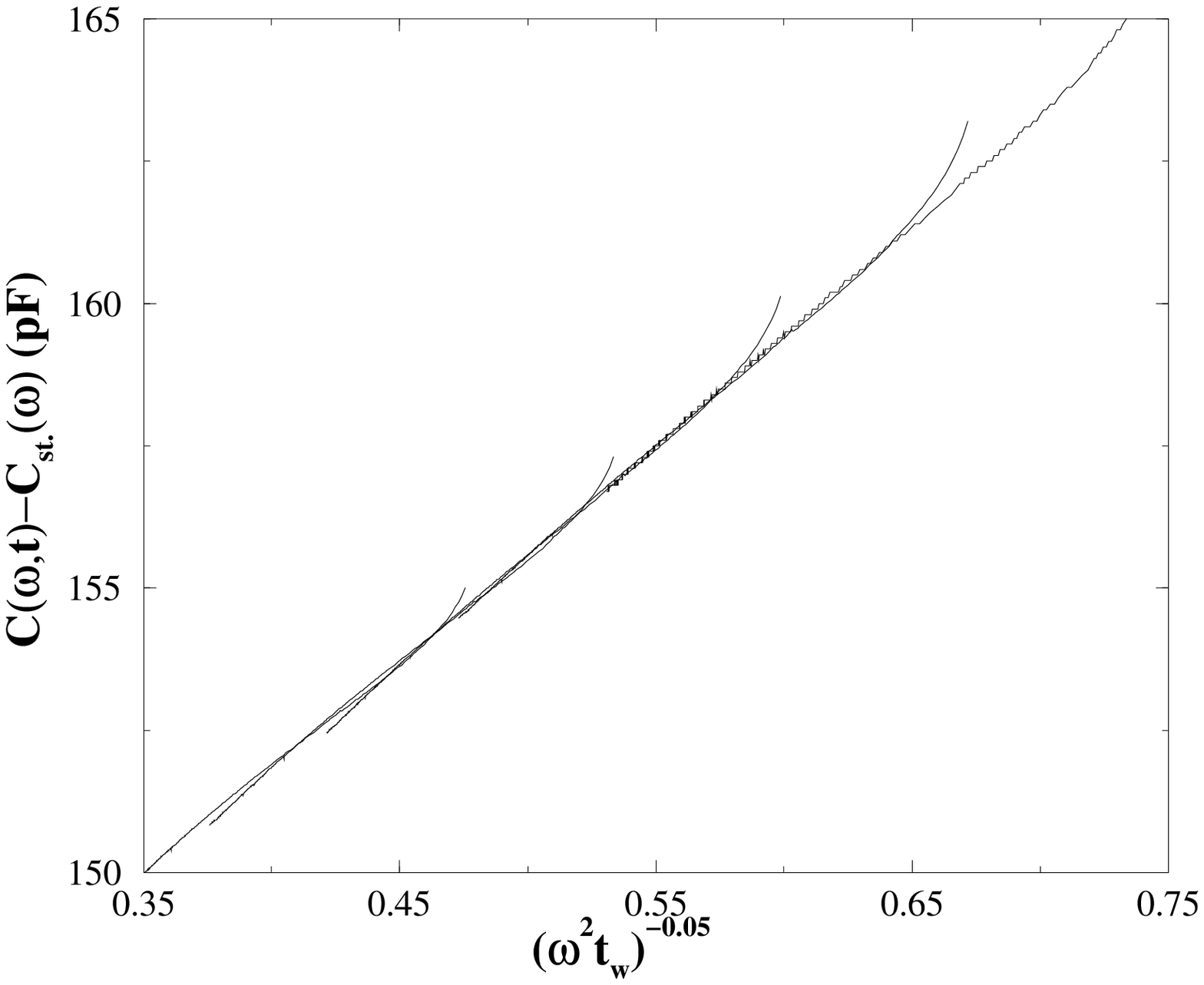}
\caption{Rescaling of the capacitance for various frequencies $\omega$ (from
$1$ kHz to $1$ MHz) and
plateau times $\tpl$. The temperature here is $\Tpl=9.8$ K. We have chosen 
the stationary part $C_{\rm st}(\omega)$
to obtain the best rescaling, as a function of $(\omega^{1/\mu} \tpl)^{-\eta}$,
with $\mu=1/2$ and $\eta=0.05$, corresponding to $\nu=0.10$.
The rescaling is not perfect for the early time part where the
time scale $t_0$ introduced in the text cannot be neglected.\label{fig2}}
\ec
\efig

\section{Frequency analysis of rejuvenation and memory}

A useful way to
parametrize the
`strength' of rejuvenation is to compare the slope ${\cal S}=\partial C'/\partial T$ of the
capacitance with temperature just after the plateau at $\Tpl$ with the
slope ${\cal S}_r$ of the reference curve, which is positive for all frequencies.
The stronger the  difference $\delta{\cal S}={\cal S}_r-{\cal S}$, the stronger the
`rejuvenation' effect.
We find again
that $\delta{\cal S}$ is a negative power-law of the frequency: small
frequencies are more efficiently `rejuvenated'. The value of the exponent is found 
to be similar
to that of $\nu$. For example, for $\Tpl=14.3$ K and $\tpl= 10 000$ s., we
find an exponent equal to $0.20$. Note that for small enough frequencies, the
capacitance increases when the temperature is decreased (${\cal S}<0$), while
the capacitance decreases (${\cal S}>0$) at high frequencies, as is
the case of the reference curve.

\bfig[t]
\bc
\epsfxsize=\linewidth
\epsfbox{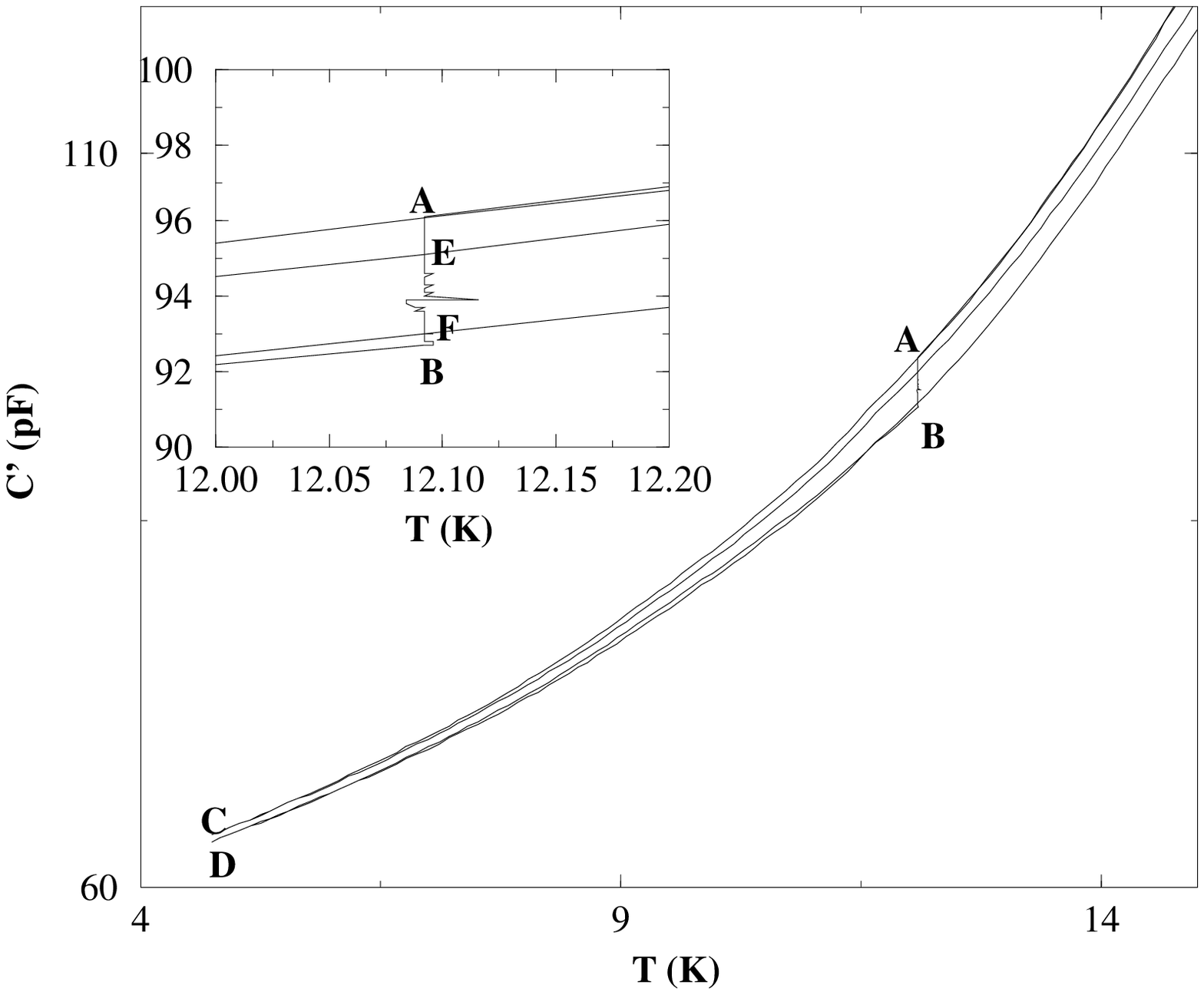}
\caption{Capacitance $C'(\omega,T)$ for two thermal histories, one without a plateau at $\Tpl$,
the other with a long plateau at $\Tpl$. The important points $A,\, B, ...,\, F$ used in the
text are defined in this figure. The inset shows a zoom of the region around 
$\Tpl$.
Note the small  temperature fluctuations during the plateau period.
\label{fig3}}
\ec
\efig

Figure 3 shows, as a function of temperature, the real part of
the capacitance $C'(\omega,T)$ for two thermal histories. The first curve is
the reference curve corresponding to a regular cooling from $T_{\max}$ to
$T_{\min}$, followed by a regular heating from $T_{\min}$ to
$T_{\max}$. The value of $C'(\omega,T_{\min})$ along this path is indicated by
point `$C$' in Figure 3, and the value of $C'(\omega,\Tpl)$ 
at $T=\Tpl$ are
point `$A$'  on the cooling curve  and point `$E$' on the heating part.

The second curve corresponds to a regular cooling from
$T_{\max}$ to $\Tpl$ (point `$A$' again), followed by isothermal aging at $\Tpl=12.1$ K
for a certain time $\tpl=20 000$ s, reaching point `$B$'. The isothermal decay
reported above is therefore $\delta C'(\omega,\tpl,\Tpl)=|A-B|$. Then regular cooling resumes
from $\Tpl$ to $T_{\min}$, where it reaches point `$D$' which is {\it below} point
`$C$'. The distance $|C-D|$ reflects some cumulative aging due to the plateau 
at $\Tpl$.
The last part of the curve is again obtained by a regular heating of the system. The value
of $C'(\omega,\Tpl)$ when passing $\Tpl$ is point `$F$'.

If the system had lost all specific memory of its stay at temperature $\Tpl$, point $F$
would be situated at point $E$ translated downwards by the cumulative part of the memory,
measured by $|C-D|$. The extra difference can thus be taken as a measure of the specific
memory of the processes taking place at $\Tpl$. We therefore define the memory indicator
${\cal M}'(\omega)$ as:
\be
{\cal M}'(\omega)=|E-F|-|C-D|.
\ee
A situation where the memory is perfectly conserved would lead to ${\cal M}'=|A-B|$.

As shown in Fig. 4, this difference is
positive in {\sc ktn}, and again behaves as a power-law with frequency.
Interestingly, the exponent describing this behaviour is very close to $\nu$.
For $\Tpl=12.1$ K and $\tpl=20 000$ s, we find a value of $0.13$, nearly
identical to the value of $\nu$ reported in Fig. 1. In the inset of Fig. 4,
we have shown the `memory ratio', i.e. the ratio of $\cal M'$ to the total isothermal decay $\delta C'$
during the
plateau studied above, for different frequencies. This ratio is around $0.5$
(showing that some memory is lost), but is nearly independent of frequencies, showing
that the frequency dependence of the memory is actually that of the aging part of
$C'$. A simple interpretation of this fact will be given below.

\bfig[t]
\bc
\epsfxsize=\linewidth
\epsfbox{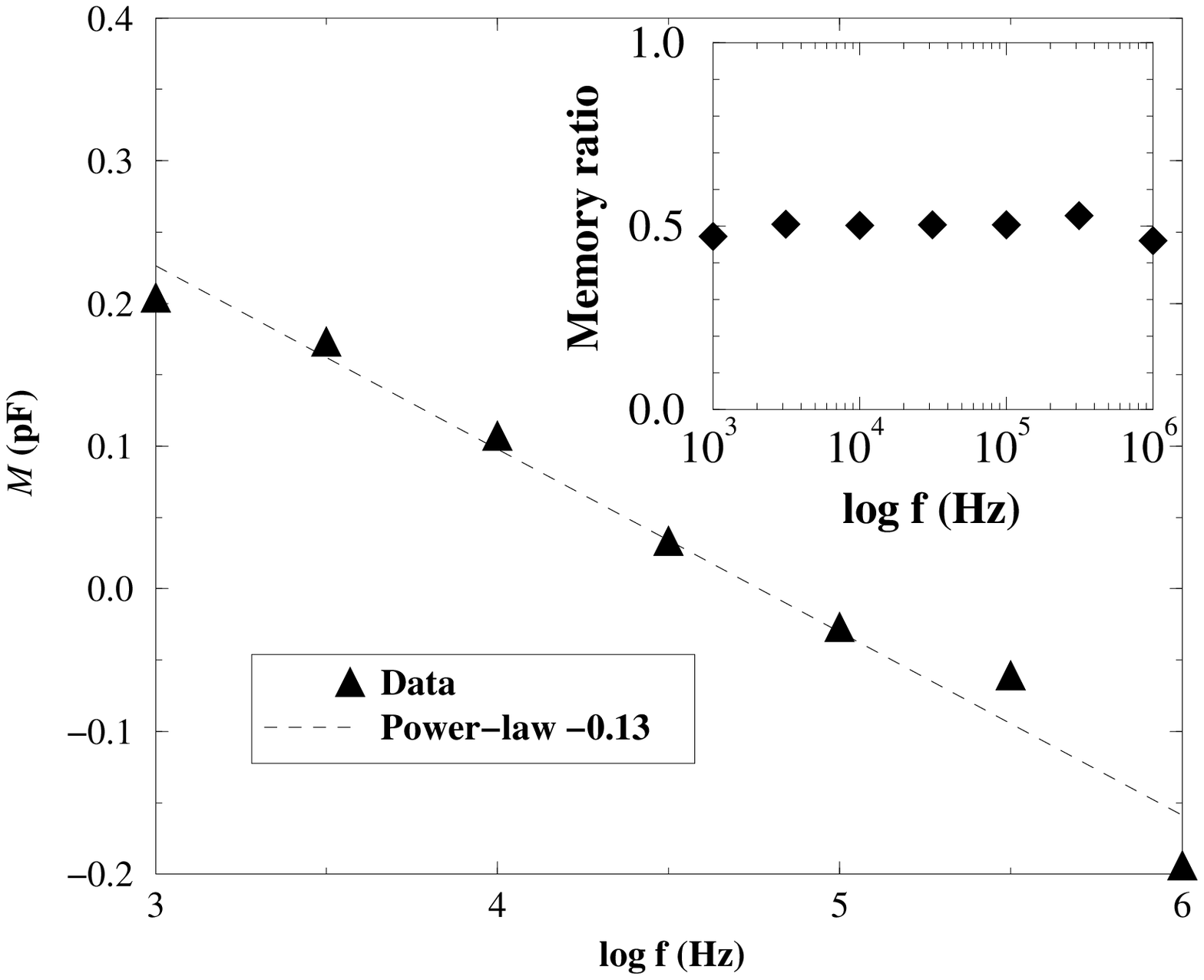}
\caption{The memory indicator ${\cal M}'(\omega)$ as a function of frequency for
$\Tpl=12.1$ K
and $\tpl=20000$s, in a log-log plot. The exponent found here is the same as in Fig. 1,
as demonstrated in the inset where the `memory ratio' ${\cal M}'/\delta C'$
is shown to
be constant, around $0.5$, over at least three decades in frequency.
\label{fig4}}
\ec
\efig

\section{Interpretation}

The microscopic origin of aging in disordered
ferromagnets or ferroelectrics is most probably the slow dynamics of domain
walls, pinned
by impurities. The important ingredient to understand the physics of these
pinned objects
is the fact that the typical pinning energy associated with a portion of wall of
linear length $\ell$ grows as $\Upsilon \ell^\theta$, where $\theta$ is a
certain exponent and $\Upsilon$
an energy scale \cite{Natterman}. This means that the time needed to equilibrate such a
portion of wall grows exponentially, as $\tau(\ell) \sim \exp(\Upsilon
\ell^\theta/T)$. Correspondingly, as
emphasized in \cite{JP}, time scales are strongly hierarchized, in the
sense that
$\tau(2\ell) \gg \tau(\ell)$. Hence, some small scale reconformations can
take place while
the overall, large scale configuration of the domain wall is fixed. This
small scale motion
between metastable states leads to better and better pinned configurations,
and therefore
a smaller susceptibility: this leads to aging, even if the overall size of
the domains does not change. The large scale coarsening leads to a progressive
increase of the average domain size $R$, and thus a decrease of the
fraction of dipoles belonging to domain walls, as $1/R$. The coexistence of
these two effects suggests that the {\it aging part} of the susceptibility
(magnetic or dielectric) should read:
\be
C_{\rm{\sc ag}}(\omega,\tpl,\Tpl) = \frac{1}{R(\tpl)} [c_{\rm st}(\omega,\Tpl) +c_{\rm
rec}(\omega,\tpl,\Tpl)],\label{theo}
\ee
where $c_{\rm st}(\omega,\Tpl)$ is the equilibrium contribution of a wall of fixed
(large) size, and $c_{\rm rec}(\omega,\tpl,\Tpl)$ is the aging contribution, due
to small scale reconformations which allow the wall to find deeper and deeper
metastable states. Simple `trap' models for these reconformations suggest that
$c_{\rm rec} \propto (\omega \tpl)^{-\nu}$ ($\nu <1$)\cite{BD}. More general (subaging)
forms are possible \cite{JP,Maass}, as observed experimentally (see Figure 2).
A possible physical mechanism is that each metastable state is visited a large number
of times. Note that in Fig. 2 we have neglected the
time dependence of $R$, which is justified if the cooling rate $|{\cal R}|$ is
small enough. In that case, the relative growth of $R$ during $\tpl$ can be
neglected.

As explained in \cite{BD,JP,Sasaki}, the reconformation contribution can account for the
rejuvenation and
memory effect. As the temperature is decreased, the small scale motion, which
was in equilibrium at the higher temperature, is suddenly driven
out of equilibrium. The aging part $c_{\rm rec} \propto (\omega \tpl)^{-\nu}$ has a
singularity for $\tpl =0$ that reflects the excess dissipation that arises
in highly out of equilibrium situations. Therefore any `micro-quench' leads to an increased
susceptibility and new aging dynamics. Simultaneously, larger scale
dynamics is effectively frozen and contributes to memory. The
distinctive feature of Eq. (\ref{theo})
is however the {\it cumulative} factor $1/R$ that always grows as
soon as the system is in its ferroelectric phase (albeit more slowly at smaller 
temperatures). This cumulative factor
is at the origin of all cooling rate effects (see the discussion in \cite{prl}). It also explains the
change of sign of the slope $\cal S$ reported above: at high frequency, the
reconformation contribution $c_{\rm rec}(\omega,\tpl,\Tpl)$ is negligible,
and the main effect is the progressive increase of $R$ which leads to a decrease
of $C_{\rm{\sc ag}}\simeq c_{\rm st}(\omega,\Tpl)/R(t)$. Therefore the slope $\cal S$
just after the plateau is positive: the capacitance decreases when the 
temperature is
lowered further. On the contrary, the rejuvenation effect is dominant at low enough frequencies,
and explains why $\cal S$ is negative. The frequency dependence of $\delta{\cal S}$ is thus
expected to be similar to that of the isothermal decay of the capacitance, as was
reported in section 4.

It is important to note that the large scale domain motion is also
responsible for the loss of memory, as suggested in \cite{12}: if, during the
period spent at lower temperatures, the domains have had enough
time to move substantially and lose track
of their previous position relative to the impurities, the memory effect is lost
More precisely, let us call $p(\tlow)$ the probability that a given wall has {\it not} substantially moved during the time
$\tlow=2(\Tpl-T_{\min})/|{\cal R}|$ spent at low temperatures, and thus retains
the memory of
the past history. The capacitance reached at $\Tpl$ on the heating curve
after a plateau at $\Tpl$ (point `F' in Fig. 3) is therefore equal to:
\bea
C_{\rm{\sc pl}}(\omega) &\simeq&
\frac{1}{R(\tpl+\tlow)} [c_{\rm st}(\omega,\Tpl)+p c_{\rm
rec}(\omega,\tpl,\Tpl)+  \nonumber
\\ 
& & + (1-p)c_{\rm rec}(\omega,0^+,\Tpl)],
\eea
where the origin of time is chosen when $\Tpl$ is first reached (point `A'). The above equation
means that a fraction $p$ of the signal has kept full
memory, while a fraction $1-p$ is totally rejuvenated. This decomposition is
only approximate,
since the memory part is affected by the time spent near $\Tpl$. The memory indicator
defined above can be written as:
\bea
{\cal M}(\omega) &=& \frac{1}{R(\tpl+\tlow)}[c_{\rm st}(\omega,\Tpl) +  \nonumber
\\
& & + c_{\rm rec}(\omega,0^+,\Tpl)]-C_{\rm{\sc pl}}(\omega),
\eea
where the first term in the right hand side is the capacitance that one would measure
in the absence of any memory ($p=0$). We
there find that the memory indicator ${\cal M}$ is related to the isothermal decay of the capacitance $\delta C(\omega,\tpl,\Tpl)$ by:
\be
{\cal M}(\omega) \simeq \frac{p(\tlow)}{R(\tpl+\tlow)}
\left[c_{\rm rec}(\omega,0^+,\Tpl)-c_{\rm
rec}(\omega,\tpl,\Tpl)\right],
\ee
or 
\be
{\cal M}(\omega)= p(\tlow) \frac{R(\tpl)}{R(\tpl+\tlow)} \delta
C(\omega,\tpl,\Tpl).
\ee
This proportionality is indeed confirmed by Fig. 4.
Note that $p$ is expected to decrease when $\tlow$ increases.
On the other hand, the ratio $R(\tpl)/R(\tpl+\tlow)$ (which is expected to
be close to $1$ in te experiments) also decreases when $\tlow$ increases.  
Therefore, the amplitude of the memory is expected to decrease as the
cooling rate decreases. We have
checked this feature experimentally, and have found that the dependence of
${\cal M}'$ on the
cooling rate ${\cal R}$ can be fitted approximately by a power law  ${\cal M}' \propto
{\cal R}^\beta$ with $\beta={0.28}$ for
$\Tpl=14.4$ K and $\tpl=10 000$ s., and ${\cal R}$ in the range $1.3$ to
$20.8$ mK/s. This
experimental dependence suggests that for very small cooling rates, the
memory effect is eventually totally erased by domain growth, as proposed in \cite{12}
for disordered ferromagnets.

\section{Conclusion}

The present experiments can be seen as a quantitative
confirmation of
the qualitative scenario of aging proposed in \cite{12,JP} for
disordered ferromagnets, which ascribes the observed effects to the motion of
pinned domain
walls. The simultaneous presence of overall domain growth and internal
reconformation modes allows one to account for the phenomenology of
temperature cycling: cumulative effects and
memory erosion are due to the former, whereas rejuvenation and memory are due
to the latter, and therefore show up for small frequencies.
We have also shown that all the quantities that measure the reconformation
contribution
behave as power-laws of frequency, and that a subaging $(\omega \tpl^\mu)$
scaling holds,
with a rather small value of the exponent $\mu \simeq 0.5$.
\vskip 1cm
{\it Acknowledgements} -- We thank S. Ziolkiewicz who has grown the {\sc ktn}
crystals used in these experiments, and J. Hammann, E. Vincent and H. Yoshino
for many useful discussions.

\end{document}